%% file: main.tex
\documentclass[sigconf,authorversion]{acmart}

\usepackage[T1]{fontenc}
\usepackage{soul}
\usepackage{xcolor}
\usepackage{todonotes}

\newcommand{\arch}{GALOIS}

\newif\ifdraft
\drafttrue
\ifdraft

\else

\fi

\AtBeginDocument{%
  \providecommand\BibTeX{{%
    \normalfont B\kern-0.5em{\scshape i\kern-0.25em b}\kern-0.8em\TeX}}}

\copyrightyear{2023} 
\acmYear{2023} 
\setcopyright{acmlicensed}
\acmPrice{15.00}
\acmDOI{10.1145/3579142.3594287}
\acmISBN{979-8-4007-0093-4/23/06}

\acmConference[BiDEDE '23]{The International Workshop on Big Data in Emergent Distributed Environments}{June 18, 2023}{Seattle, WA, USA}
\acmBooktitle{The International Workshop on Big Data in Emergent Distributed Environments (BiDEDE '23), June 18, 2023, Seattle, WA, USA}
\begin{document}

\title{GALOIS: A Hybrid and Platform-Agnostic\\ Stream Processing Architecture}

\author{Tarek Stolz}
\email{tarek.stolz@rwth-aachen.de}
\orcid{0009-0000-5735-4318}
\affiliation{%
  \institution{RWTH Aachen University}
  \streetaddress{Ahornstrasse 55}
  \city{Aachen}
  \country{Germany}
  \postcode{52074}
}

\author{István Koren}
\email{koren@pads.rwth-aachen.de}
\orcid{0000-0003-1350-6732}
\affiliation{%
  \institution{Chair of Process and Data Science,\\ RWTH Aachen University}
  \streetaddress{Ahornstrasse 55}
  \city{Aachen}
  \country{Germany}
  \postcode{52074}
}

\author{\href{https://orcid.org/0000-0003-0049-8112}{Liam Tirpitz}, \href{https://orcid.org/0000-0002-8970-6282}{Sandra Geisler}}
\email{{tirpitz,geisler}@cs.rwth-aachen.de}
\orcid{0000-0003-0049-8112}
\affiliation{%
  \institution{Data Stream Management \& Analysis,\\ RWTH Aachen University}
  \streetaddress{Ahornstrasse 55}
  \city{Aachen}
  \country{Germany}
  \postcode{52074}
}

\renewcommand{\shortauthors}{Stolz et al.}

\begin{abstract}
With the increasing prevalence of IoT environments, the demand for processing massive distributed data streams has become a critical challenge.
Data Stream Processing on the Edge (DSPoE) systems have emerged as a solution to address this challenge, but they often struggle to cope with the heterogeneity of hardware and platforms.
To address this issue, we propose a new hybrid DSPoE architecture named \arch, which is based on WebAssembly (Wasm) and is hardware-, platform-, and language-agnostic.
\arch~employs a multi-layered approach that combines P2P and master-worker concepts for communication between components.
We present experimental results showing that operators executed in Wasm outperform those in Docker in terms of energy and CPU consumption, making it a promising 
option for streaming operators in DSPoE.
We therefore expect Wasm-based solutions to significantly improve the performance and resilience of DSPoE systems.

\end{abstract}




\settopmatter{printfolios=true}
\maketitle

\input{mainmatter/1-introduction.tex}
\input{mainmatter/2-requirements.tex}
\input{mainmatter/3-concept.tex}
\input{mainmatter/4-prototype.tex}
\input{mainmatter/5-evaluation.tex}
\input{mainmatter/6-related-work.tex}

\input{mainmatter/7-conclusion.tex}

\begin{acks}
Funded by the Deutsche Forschungsgemeinschaft (DFG, German Research Foundation) under Germany's Excellence Strategy - EXC-2023 Internet of Production - 390621612.
\end{acks}
\bibliographystyle{ACM-Reference-Format}
\bibliography{bidede}

\end{document}

%% file: mainmatter/1-introduction.tex
\section{Introduction}
\label{sec:intro}

Modern manufacturing environments produce tremendous volumes of data at very high frequencies.
For example, in fine blanking (an industrial stamping process) a single production line can produce data up to 6 GBit/s \cite{glebke2019case}.
In our vision of an \textit{Internet of Production}~\cite{BDJ*22}, manifested by a large research cluster of excellence, the processing and 
sharing of high-frequency machine data across organizations is fundamental to enable large-scale near real-time analysis and fuel \textit{digital shadows} to facilitate smart manufacturing applications.
However, the high degree of heterogeneity of machines, programmable logic controllers, and network connectivity creates a challenging environment for data processing.
In addition, the increased number of variants within production lines and ultimately also the need to react quickly to incidents in light of sustainability and resilience efforts demand fast and adaptive data stream management solutions.

Mature distributed stream processing (DSP) architectures, such as Apache Storm or Twitter Heron, are available for several years now and have proven to handle high amounts of streaming data. 
They were designed to exploit the rich resources of nodes in data centers on-site or in the cloud, distributing the workload to homogeneously equipped servers. 
Transmitting the data of multiple production lines to a central data center leads to network congestion and intolerable latencies for near real-time applications.
Edge processing architectures move parts of the data processing (operators) to the edge and the fog, close to the data sources~\cite{cao2020overview,varghese2016challenges}.
For edge environments, many challenges arise which are opposed to those in the cloud ~\cite{varghese2016challenges}.
For example, insecure, unstable, limited network connections and failing or moving devices create a highly dynamic environment for data processing.
Further, edge devices have a high degree of heterogeneity in, e.g., hardware or operating systems, and often have limited resources and battery life.
Distributed stream processing on the edge (DSPoE) systems are designed to handle these issues.
However, operators in DSPoEs are often implemented language- or hardware-dependent which makes these systems inflexible as well as hard to maintain and extend.
Additionally, DSPoEs usually rely on a centralized management component leading to high communication overhead.

To address these challenges, we introduce \arch, a hybrid and platform-agnostic stream processing architecture.
Its architecture is multi-layered to provide flexibility and distribute workload efficiently, employing centralized as well as decentralized management depending on the layer.
Re-optimizations of the query execution plan are propagated from lower to higher levels to reduce as much communication overhead as possible with a central, overarching master node.
Furthermore, we use WebAssembly (Wasm), a new byte code format that can be run on diverse hardware, to implement operators in a hardware-, platform-, and language-independent manner, providing the foundation of \arch, which we will discuss in more detail in the following.

The paper is organized as follows.
In Section \ref{sec:requirements}, firstly, the requirements for a hybrid and hardware-agnostic DSPoE are analyzed.
Based on these requirements, we present the design of the \arch~architecture in Section~\ref{sec:concept}. 
In Section~\ref{sec:prototype} a first prototype implementing this concept is described, which is evaluated in Section~\ref{sec:evaluation}. 
Finally, in Section~\ref{sec:relatedwork}, we discuss the presented architecture in light of the current state of the art and draw conclusions in Section~\ref{sec:conclusion}.

%% file: mainmatter/2-requirements.tex
\section{Requirements}
\label{sec:requirements}
In the following, we outline the key requirements that a DSPoE system should meet to effectively operate within highly dynamic environments, accommodating massive data streams and heterogeneous devices at the edge.
These requirements have been derived through a thorough analysis of edge environment characteristics in manufacturing, as well as by examining the necessary attributes and addressing unresolved challenges in existing DSPoE systems.

\textit{Platform-agnostic operator execution} \textbf{(R1)}: To accommodate the diversity of edge environments, stream operators should be designed for execution irrespective of hardware, language, and platform constraints.\\
\textit{Native edge streaming and processing} \textbf{(R2)}: To enable fast response times, the data should be streamed directly point-to-point between processing edge devices without any intermediaries.
Furthermore, the execution of operators should take place on the edge level.\\
\textit{Network-aware management and administration} \textbf{(R3)}: The edge network's topology and properties, such as link quality and device distance, should be considered during query processing to minimize network-induced latencies and backpressure.\\
\textit{Global logical optimization} \textbf{(R4)}: Logical query optimization should be done on a global level based on heuristics, as it will influence the number and types of operators which are subsequently be distributed to edge nodes in operator placement.\\
\textit{Local physical optimization} \textbf{(R5)}: To leverage the latest, contextually relevant information for query optimization and decrease communication overhead, physical query optimization should occur at the same level as operator execution, thereby enhancing layer autonomy and modularity.\\
\textit{Spatial organization in cluster units} \textbf{(R6)}: The system should take into account spatial constraints when deploying to and managing edge devices. 
The system should prefer to deploy adjacent operators in dense accumulation points of edge devices (cluster units) to establish the shortest possible communication paths.\\
\textit{Decentralized query reconfiguration} \textbf{(R7)}: If query plan adjustments are necessary, these should be carried out in a decentralized and independent manner without pausing the complete system.\\
\textit{Administrative escalation model} \textbf{(R8)}: Higher-level administrative units in the architecture should only intervene when re-optimization at lower levels is unfeasible.\\
\textit{Preventive fault tolerance} \textbf{(R9)}: 
Resilient fault tolerance is important to elevate the robustness of a query.
In the event of a client failure, a query should be able to deliver results. 
This can be achieved by increasing the robustness of a query through more selective operator placement and providing multiple paths between operators~\cite{frontier,R-MStorm}.
These core requirements inform the design of our architecture.

\paragraph{Resulting design decisions}
As edge environments can be highly dynamic, the metrics used for operator placement in DSPoEs may already be outdated when the query plan has been deployed~\cite{Neb,DART}. 
Hence, the development of DSPoE systems is tending more and more towards a decentralized deployment of queries and functionalities \cite{tiot,Neb}. 
As a result, Pinchao et al.~\cite{DART} organize edge devices within a P2P network and perform query optimization not at deployment time but at runtime. 
Because the query is optimized incrementally, there is a lower risk that query execution is no longer optimal as time progresses \textbf{(R7)}. 
However, many P2P networks abstract away from the underlying network topology, which can lead to non-optimal operator placement~\cite{manet-p2p,p2p-performance}. 
In order to achieve low latencies, it is important that operators are positioned as close as possible to the sources and adjacent operators are also positioned in near vicinity to each other.
Therefore, \emph{cluster-based methods} for organizing the underlying physical devices on the lowest level seem to be a viable solution to this problem \textbf{(R6, R3)}~\cite{manet-p2p,Mobi,MS-DSPMC,R-MStorm,F-MStorm}.\\
In this context, Mobile Ad-hoc Networks (MANETs) are attractive, as they enable cluster computing on the edge~\cite{manet-p2p}.
MobiStream~\cite{Mobi} defines fixed locations for clusters in MANETs, but does not accommodate dynamically forming, disappearing, and reorganizing clusters. 
For this reason, clusters should support mobility and be able to be registered autonomously \textbf{(R6)}.
For \arch~, we opted to organize spatially separated clusters in regions for scalability and flexibility, similar to Mobile Storm's concept~\cite{MS-DSPMC} \textbf{(R6)}.
Furthermore, additional regions can be added and thus a spatial scaling as in SpanEdge is enabled \textbf{(R4,R8)}~\cite{spanEdge}. 
Cluster-to-cluster communication and region-to-region communication overhead should be kept as low as possible and implemented in a 
P2P fashion to provide high flexibility and autonomy. 
Therefore, query operator execution should be done on cluster and edge device level \textbf{(R2)}~\cite{manet-p2p}.\\
During query runtime, continuous query optimization shall be done in the cluster which should be as autonomous as possible and only be escalated upwards to managing nodes if necessary (e.g., due to unresolvable overload situations) to reduce the administrative overhead~\textbf{(R8)}~\cite{Neb}.
The query's physical and logical optimization should be divided between local and global layers, ensuring global instances are not burdened with complete knowledge of the underlying edge devices and operators.
Conversely, physical optimization, heavily reliant on dynamic metrics, should be performed as close as possible to the execution site \textbf{(R6,R7}).\\
To maximize fault-tolerance, \arch~ will use a probability model superior to checkpoint- or replicant-based schemes~\cite{R-MStorm} and select only particularly fail-safe nodes in the operator placement \textbf{(R9)}~\cite{frontier} making the architecture more robust.\\
Further, the architecture is designed to be extensible in terms of different query languages similar to Calcite or Governor~\cite{calcite,governor}.
Finally, to execute operators on a heterogeneous system landscape, operators should be executable independent of hardware, platform, and language. 
All DSPoE we are aware of to date are language-dependent (\textbf{R1})~\cite{grizzly,DART,MS-DSPMC}, which we address with the architecture of \arch.

%% file: mainmatter/3-concept.tex
\section{Concept}
\label{sec:concept}
Based on the requirements and design decisions explained in the previous section, we conceptualized a system architecture, which is described in detail in the following.
\begin{figure}[tbp]
    \centering
    \includegraphics[width=0.49\textwidth]{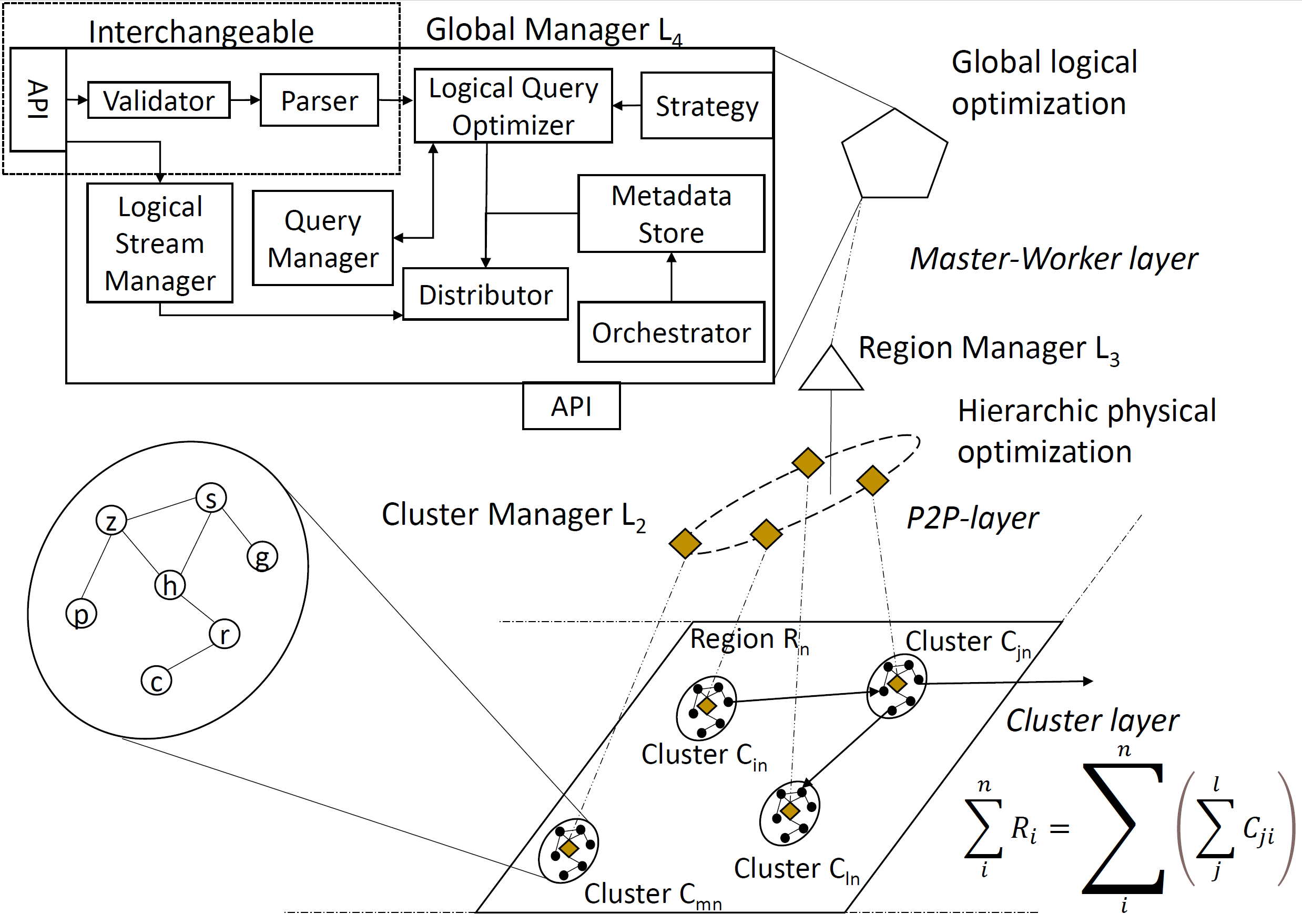}
    \caption{The \arch~architecture}
    \vspace*{-0.5cm}
    \label{fig:arch}
\end{figure}

Figure \ref{fig:arch} shows an overview of the various components in \arch. 
The system is hierarchically structured, employing four levels of communication~\cite{Neb} \textbf{(R6)}. 
The lowest level is the \emph{Cluster Level} where nodes are organized in clusters and transfer tuples in data streams. 
Clusters represent accumulation points of nodes that have a physical proximity which is exploited \textbf{(R2)}.
Within the cluster, the data streams are exchanged between the nodes.
In Figure~\ref{fig:arch}, the clusters are represented as circles and network connections as solid lines between the nodes, which are shown as black dots in the clusters.
Cluster-to-cluster data streaming is depicted as solid lines between the clusters and is performed via gateways. 
The yellow squares represent the \emph{Cluster Managers} (CM) which are interconnected by a P2P network (shown as a dashed ellipse) and, if necessary, are communicating with the \emph{Region Manager} (RM) of their region (shown as a triangle). 
To increase scalability and build local, independent subsystems, each cluster is assigned to a RM. 
Regions can be defined based on the needs of the application, e.g., by grouping geographically close clusters.
The RM and GM are organized in a master (GM) worker (RM) layer.

The \emph{Global Manager} (GM, depicted as pentagon) is primarily responsible for three tasks: (1) registration of logical data sources (schema), (2) receipt of the logical query plan, and (3) logical query optimization \textbf{(R4)}. 
The registration of a schema can be done in two ways: (1) either a human user registers the schema and the corresponding data sources or (2) the data sources act autonomously through corresponding interfaces and register themselves and their schema. 
There are several formal query languages that can be used to register and describe a query. 
Similar to Apache Calcite or NebulaStream the GM should be able to support multiple languages~\cite{calcite,Neb}.
The optimizer in the GM can optimize the logical query plan utilizing an optimization strategy. 
Unlike other systems, no physical optimization is performed in the GM and a strict separation between logical and physical optimization is enforced. 
A successfully deployed query is stored in the \emph{Query Manager} of the GM to enable multi-query optimization. 
Furthermore, the \emph{Query Manager} can provide the user with information about the currently running queries and their state. 
The optimization strategy should be easily interchangeable to allow different optimization strategies, comparable to the Governor system~\cite{governor}. 
After the query has been optimized, it is divided into subqueries and sent to the \emph{Region Managers} (RMs) by the \emph{Distributor}. 
In addition to the subqueries, information important for interregional communication is provided to the RMs to enable interregional data transfer between nodes in different regions. 
The splitting into subqueries is done using metadata which is stored in the \emph{Metadata Store}. 
These comprise region-specific metadata, e.g., the source nodes in a region, region performance statistics, such as free resource capacities, running subqueries, or key metrics such as the average tuple throughput. 
The \emph{Orchestrator} monitors the region topology and the start of new regions, which can be dynamically added or deleted.\\
\emph{Region Managers (RMs)} are introduced as the next layer of abstraction and enable scalability and locality, by independently managing groups of clusters. 
A region can be specified geographically or on the basis of latency zones, for example, in order to identify the most latency-optimal regions, similar to SpanEdge~\cite{spanEdge} \textbf{(R3)}.
It is assumed that latency for geographically distant points is significantly higher than for spatially closer nodes. 
The notion of proximity can be defined on the implementation level and depends on the application.
Therefore, it is important that the subqueries are chosen such that there is only as much cross-region communication as necessary.
The core task of the RM is to distribute the subqueries to the \emph{Cluster Manager} based on summary statistics, e.g., cluster-specific operator throughput or average queue size.
The RM distributes the logical subqueries to the \emph{Cluster Manager} accordingly and manages the assignment in the local query manager for possible reconfigurations \textbf{(R7)}. 
In addition, the RM also serves as a bootstrap node for clusters that appear in the region providing the new \emph{Cluster Manager} with all necessary information to connect to the P2P network.\\
The \emph{Cluster Managers} (CM) are organized in a P2P network. 
The most important tasks of the CM are operator placement, collection of metadata, monitoring of the GALOIS clients, and communication with other CM within a P2P network to distribute workloads at runtime. 
Based on the gathered metrics, subquery reoptimization is organized autonomously by the CMs within the P2P network and thereby metadata exchange is reduced \textbf{(R8)}.
Furthermore, the P2P network should ensure that new clusters can be set up mobile (e.g., for vehicles or smartphones) and dynamically over time. 
Node failures in the cluster should not stop the processing of the entire query. 
Thus, delays like in MobiStream (stop the hole query on failure) have to be avoided~\cite{Mobi}.
At the lowest level of abstraction GALOIS clients are located consisting of edge devices with the DSPoE client installed. 
Each node is assigned to a cluster by node discovery. 
If there is no CM nearby, an isolated node can take over the CM role building a new cluster. 
The new cluster remains inactive until a sufficiently large number of nodes are registered in the cluster (the cluster size is configurable to reduce management overhead). 

Beside being the CM (h), nodes can take six additional roles in a cluster: Sleeper (s), Producer (p), Consumer (c), Replica (r), Gateway (g), and Sink (z). 
This role scheme allows for probabilistic models which assign roles according to reliability for fault tolerance \cite{R-MStorm}. 
Nodes which are not the CM are initialized as \textit{s} nodes. 
Sleepers do not yet have a differentiated role and are available to the CM as a resource. 
As soon as the operators of a query are distributed corresponding nodes get the producer (\textit{p}) role assigned and which are the source of a data stream. 
All subsequent operators in the subquery are consumers (\textit{c}) which are processing the data stream incrementally. 
To implement a fault tolerance scheme nodes get the role \textit{r} (replica) assigned, 
which are statistically particularly fail-safe. 
These nodes regularly check the \textit{c} node assigned to them and take over its functionality in the event of a failure. 
In order to transfer data between clusters, cluster-to-cluster communication is required; this functionality is provided by \textit{g} nodes. 
They act as gateways between the clusters. 
The \textit{z} token is assigned to sink nodes by the CM, which provide APIs for data consumption by users or applications. 
These node types and the general architecture are realized in a prototypical implementation.

%% file: mainmatter/4-prototype.tex
\section{Prototype}
\label{sec:prototype}
In order to realize the requirement of platform independence in the prototype, a form of abstraction is needed. 
Common solutions to enable platform independence are virtual machines or Docker containers.
Since the stream operators should be executed on edge devices, we target an execution environment requiring minimal resources and startup effort. 
With the binary instruction format WebAssembly (Wasm)\footnote{\url{https://webassembly.org}}, especially suited for serverless scenarios~\cite{Gackstatter},  
operators can be executed as functions within the Wasm runtime environment. 
In literature, Wasm is regarded as a potential alternative to Docker, especially in edge scenarios, as it proved to be more memory-efficient 
than Docker~\cite{Gackstatter,Hampau}. 
Hence, we implement the stream operators in our system targeting Wasm to make their execution platform-, language-, and hardware-agnostic \textbf{(R1)}.
We chose Rust as the implementation language also for the client and the managers, due to its native Wasm support and integration, as well as its ability to produce small binaries.
However, any other language could be chosen for the operators and the choice is independent of the client and managers' implementations.

\begin{figure}[ht]
    \centering
    \includegraphics[width=0.48\textwidth]{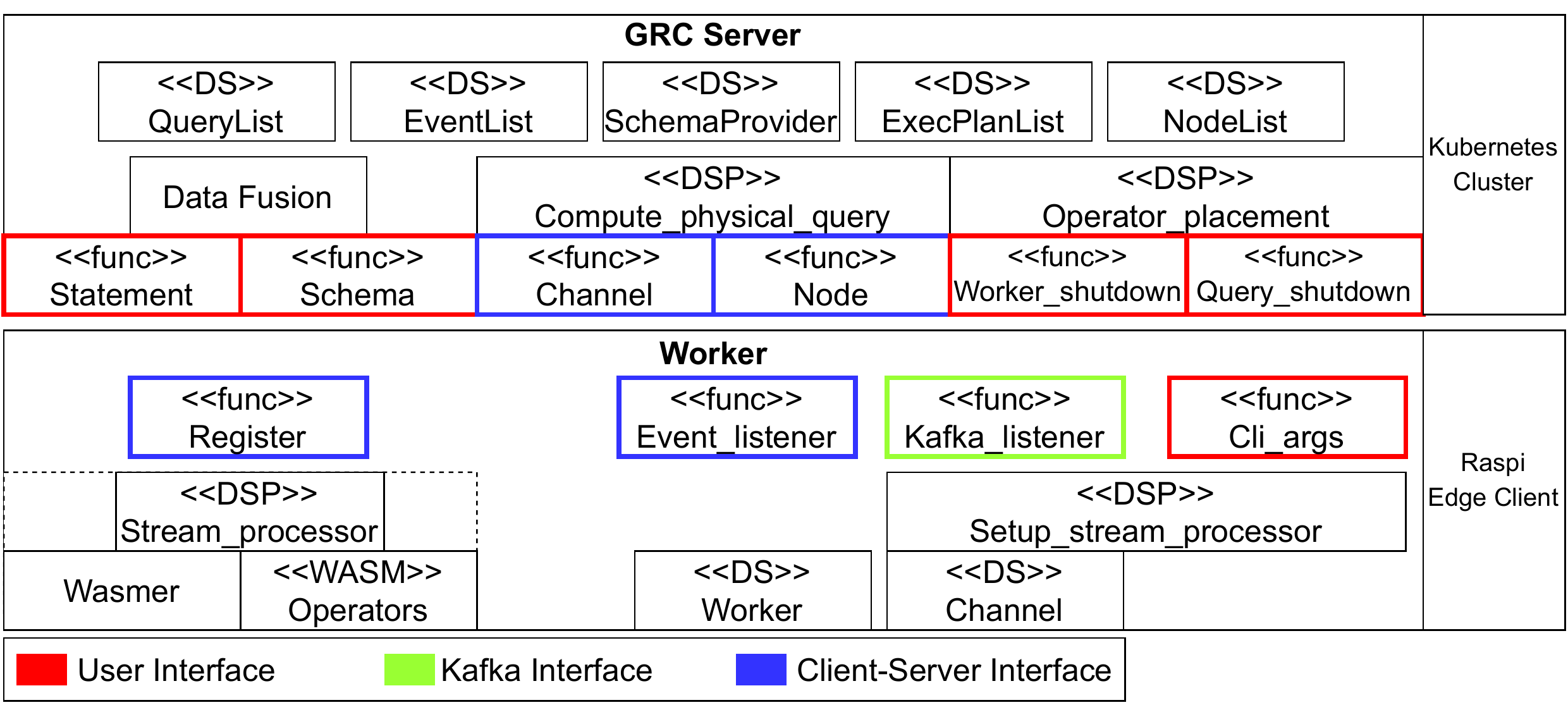}
    \caption{Overview of the prototype and its components}
    \label{fig:overview}
    \vspace{-0.3cm}
\end{figure}

The implementation of the first prototype followed a bottom up approach, concentrating on the development of operators in Wasm as a proof-of-concept. 
To test the operator execution, only one cluster with multiple nodes is created. 
Data transfer between nodes is implemented using Kafka, but will be replaced in the next step by efficient inter-node communication.
In the prototype, a single server component represents GM, RM, and CM in unity (GRC server). 
For scaling, the components can be separated later. 
Currently, the GRC server comprises modules for query, schema, and node management, as well as physical query plan creation and a simple operator placement \textbf{(R5)}.
Additionally, Data Fusion is integrated to execute logical query optimization.
The components will be complemented in the future with additional features, such as cluster and resource management and fault tolerance protocols including backpressure mechanisms suited for edge environments \textbf{(R9)}.
The GRC server communicates with the nodes via server-sent events (SSE) including the \textit{nodeID} and the corresponding command. 
Commands are also used to deploy operators to nodes after the GRC server has completed the operator placement.
The GRC server is deployed as a Docker container in a Kubernetes cluster and the workers run on Raspberry Pis \textit{model B Rev 1.5}.
Figure~\ref{fig:overview} provides an overview of the most important components. 

%% file: mainmatter/5-evaluation.tex
\section{Evaluation}
\label{sec:evaluation}
\begin{figure}[htbp]
    \centering
    \includegraphics[width=0.4\textwidth]{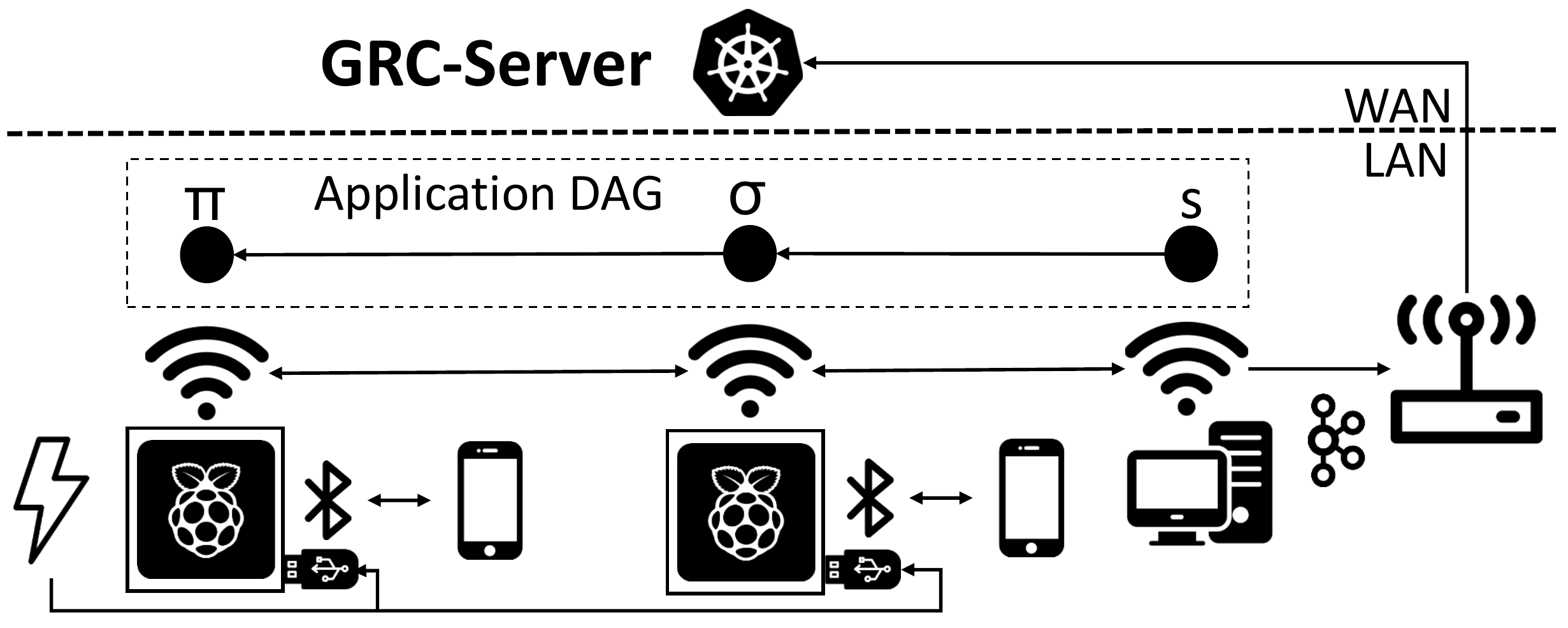}
    \caption{Evaluation Setup}
    \vspace*{-2ex}
    \label{fig:setup}
\end{figure}

In the preliminary phase of our research, we conducted an assessment of the prototype system, with a primary focus on comparing the performance of implementing operators in Wasm to that of Docker containerization.
This evaluation aimed to explore the differences in latency, energy consumption, throughput, and CPU load between the two approaches.
Our hypothesis posited that Wasm, due to its near-native exploitation of hardware characteristics and reduced initialization overhead, would exhibit superior efficiency in these performance metrics.
The evaluation utilized the ELEC dataset~\cite{elec}, comprising 45,312 tuples on half-hourly electricity prices from New South Wales, Australia.
It covers a period of 942 days and contains the following attributes: Date, Day, Period, NSWprice, NSWdemand, VICprice, VICdemand, transfer. 
The evaluation setup is shown in Figure~\ref{fig:setup}. 

The setup comprises two Raspberry Pis 4, Model B Rev 1.5, running Linux raspberrypi 5.15.76-v8+ on 64-bit basis.
They represent the edge devices in one cluster to which operators can be distributed.
Besides the Raspberry Pis, a Windows 10 computer was used, which acted as the producer, creating a data stream from the aforementioned dataset, running a local, dockerized Kafka instance.
Further two DollaTek UM25C USB power meters were used to record the power consumption of the Raspberry Pis.
The power meter were connected to two separate smartphones as the monitoring software only allows one Bluetooth connection to a power meter at a time. 
The GRC server was deployed to the Kubernetes infrastructure of the Internet of Production (IoP) located in a different network.
A continuous query, including a projection onto the date and period attributes and 18 selection conditions (to increase query complexity), was registered at the GRC server, which distributed the selections to one Raspberry Pi and the projection to the second Raspberry Pi to measure the processing costs for each operator individually.
We simulated varying system load, by generating tuples with different data rates at the producer, specifically 125 T/s (Tuples/second), 500 T/s, and 1000 T/s. 
 For each input rate, power consumption and CPU load were measured at the edge devices, as well as throughput (T/s at each operator and for the complete query), and query latency (time of a tuple between entering and leaving the system).

\subsection{Results}
\label{ch:results}
\begin{figure}[t]
    \centering
    \includegraphics[width=0.38\textwidth]{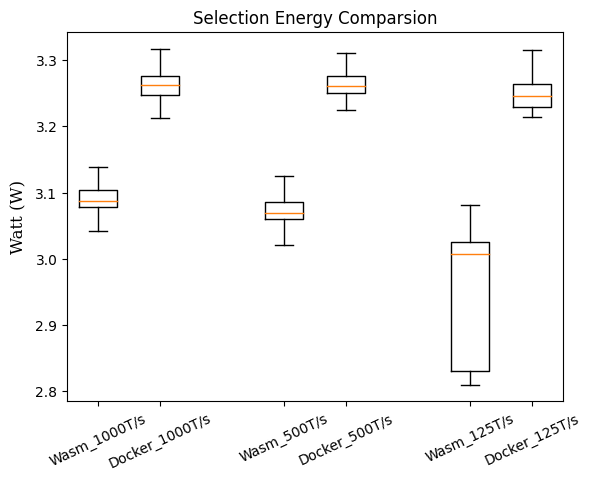}
    \caption{Energy Consumption Comparison}
    \label{fig:energy-comp-boxplot}
    \vspace{-0.5cm}
\end{figure}

We present the results of the described experiments in the following, where percentages indicate average values.
\textit{}{Operator Processing Time}
The Operator processing time (OPT) is the difference between the timestamp directly before the execution of the operator and the timestamp afterwards. 
Our experiments show, that for selection Wasm outperforms Docker for each of the input rates with 9\% less OPT.
\paragraph{Power Consumption}
Energy consumption is recorded in Watts per second by the power meters at operator execution. 
Figure~\ref{fig:energy-comp-boxplot} contains six boxplots showing the power consumption of the selection execution. 
On the x-axis the boxplots are labeled with the experiments - each combination of execution environment (Wasm or Docker) and input rate. 

\begin{figure}[t]
    \centering
    \includegraphics[width=0.38\textwidth]{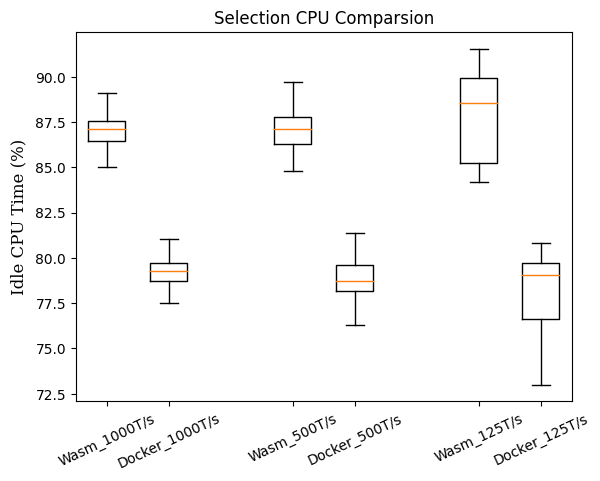}
    \caption{Idle CPU Time Comparison}
    \vspace*{-2ex}
    \label{fig:cpu-comp-boxplot}
    \vspace{-0.3cm}
\end{figure}

The energy consumption for the Docker experiment was 5.72\% higher than for Wasm at the 1000 T/s, 6.56\% percent higher than for Wasm at 500 T/s, and 10.09\% percent higher than for Wasm at the 125 T/s input rate.
\textit{CPU Load}
The average CPU load was recorded as percentage of CPU Idle Time (CIT) during tuple processing only, where higher values indicate a lower CPU load and vice versa. 
The CPU load is considered as an indicator for the potential additional workload for Docker virtualization.
Figure~\ref{fig:cpu-comp-boxplot} shows six boxplots also labeled on the x-axis with the combination of execution environment and input rate.
On the y-axis the percentage of CIT during selection execution is plotted. 
The CIT for the Wasm experiment was 11.36\% percent higher than for Docker at 1000 T/s, 18.89\% percent higher than for Docker at 500 T/s, and 23.05\% percent higher than for Docker at the 125 T/s input rate.
The projection showed no significant differences, only the standard deviation was smaller for all measurements.

\subsection{Discussion}
\label{ch:discussion}
Our results support the assumption that Wasm's processing time per tuple is more efficient than Docker's due to its lower CPU intensity.
This efficiency stems from several factors: Docker runs as a standalone process, managing its environment and virtual network, while Wasm, embedded in Rust, only executes operators.
Despite expecting a more significant difference in selection, the OPT remains relatively constant, possibly due to operational and CPU limitations.
Notably, Wasm's OPT exhibits less fluctuation for both operators compared to Docker, likely due to the latter's additional CPU overhead.

As expected, Docker consumes more power due to containerization overhead.
Interestingly, the CPU and energy consumption difference between Docker and Wasm decreases at higher input rates, with the largest gap at 125 T/s.
Further experiments are needed to investigate the relationship between these differences and input rates.
Our results align with Hampau et al.~\cite{Hampau}, as Wasm consumes less energy than Docker.
The highest difference at 125 T/s might result from Wasm's better OPT and increased waiting time for tuples, which vanishes at higher input rates due to backpressure.

The OPT, limited by CPU, could be reduced with multi-threading but at the cost of increased energy consumption, presenting a tradeoff~\cite{Neb}.
Our experiments aimed to evaluate Wasm's feasibility as a streaming operator library.
Results indicate Wasm as a potential lightweight, hardware-independent FaaS alternative to Docker for DSPoE.
We plan to extend the operator library, including standard operators like joins and sophisticated features like machine learning.
In-depth experiments will assess complex queries, operator placement, query reoptimization, and communication costs.
This will provide a comprehensive assessment of the proposed operator network's effectiveness and benefits for data streaming.

%% file: mainmatter/6-related-work.tex
\section{Related Work}
\label{sec:relatedwork}
The DSPoE systems NebulaStream, MobiStream, DART, Apache Nifi, and MobileStorm are related systems that enable stream processing employing concepts from edge computing~\cite{Neb,Mobi,DART,PIE,MS-DSPMC} and we analyzed them according our posed requirements.
Many of these systems, such as MobileStorm or Apache Nifi, use a central administrative component, e.g., responsible for monitoring, query distribution, client management, and fault handling, while the clients only execute the operators. 
In contrast, DART uses a purely decentralized approach and offloads many of the functions to the individual clients organized in a P2P network.
In \arch, we employ a hybrid model to combine the best of both approaches, offering high scalability and flexibility.
Almost all of the above systems use the Java Virtual Machine (JVM) as execution environment for platform- and hardware-agnostic operator execution. 
However, the JVM is not language-independent.
NebulaStream uses query compilation, i.e., queries are distributed to the edge devices as generic query plans, where it is converted into a C++,  compiled, and executed~\cite{grizzly}. 
None of the discussed approaches are language-independent.
Instead of relying on the JVM, the online Wasm compiler presented by Groppe and Reimer~\cite{groppe19code} contributes towards language-independent query processing in networks of internet browsers, but does not focus on data stream processing.

Cloud-based Data Stream Processors (CDSP) have been widely used and adapted as the basis for DSPoE systems in the past. 
However, most DSPoE systems in this area were developed in Java lacking hardware- or language-independent execution of operators.
CDSPs also miss a concept for fault handling that fits the dynamics of the edge environment and they usually do not allow runtime optimization of queries, since the performance in the cloud is assumed to be stable. 
Due to the contradictory requirements of cloud and edge, it makes sense to develop a DSPoE framework which does not to build on an existing CDSP.

%% file: mainmatter/7-conclusion.tex
\section{Conclusion and Outlook}
\label{sec:conclusion}
In this paper we present a new concept for a hybrid multi-layer DSPoE architecture \arch. 
Its architecture facilitates cluster communication on the edge using a P2P network concept enabling re-optimizations on a local level and thereby reducing communication overhead. 
Also, nodes in a cluster can take different roles, which will guide the operator placement.
At the same time GALOIS offers scalability by introducing a region management layer containing several clusters and a global manager implemented in a master-worker fashion.
Opposed to existing DSPoEs, \arch~is hardware-, platform, and language-agnostic by using WebAssembly as execution environment for streaming operators distributed to the edge.
We showed in experiments on a prototypical implementation of our architecture, that operator execution in Wasm is more resource-efficient than in Docker in terms of Operator Processing Time and CPU and energy consumption. 
The lower throughput observed in our experiments can be attributed to the conservative use of the CPU, but this issue could be resolved in the future by leveraging higher parallelism with multi-threading.
To summarize, Wasm represents a viable alternative to Docker for executing operators in DSPoE environments.
However, since Wasm is still in development, it has some inherent limitations that need to be considered.
For example, it is currently not possible to pass Wasm complex data objects without additional glue code.
We plan to extend \arch~by (1) expanding the Wasm stream operator library to allow for more complex queries, (2) integrate a flexible QoS and data quality monitoring to enable a dynamic re-optimization of queries, (3) implement a full-fledged prototype including all hierarchy levels based on the envisioned communication infrastructure also considering in-network processing.